\newfam\scrfam
\batchmode\font\tenscr=rsfs10 \errorstopmode
\ifx\tenscr\nullfont
        \message{rsfs script font not available. Replacing with calligraphic.}
        \def\scr{\cal}

\else   
        \font\sevenscr=rsfs7
        \font\fivescr=rsfs5
        \skewchar\tenscr='177 \skewchar\sevenscr='177 \skewchar\fivescr='177
        \textfont\scrfam=\tenscr \scriptfont\scrfam=\sevenscr
        \scriptscriptfont\scrfam=\fivescr
        \def\scr{\fam\scrfam}
        \def\cal{\scr}
\fi
\newfam\msbfam
\batchmode\font\twelvemsb=msbm10 scaled\magstep1 \errorstopmode
\ifx\twelvemsb\nullfont\def\Bbb{\bf}
        
	\font\eightbbb=cmb10 at 8pt
	\message{Blackboard bold not available. Replacing with boldface.}
\else   \catcode`\@=11
        \font\tenmsb=msbm10 \font\sevenmsb=msbm7 \font\fivemsb=msbm5
        \textfont\msbfam=\tenmsb
        \scriptfont\msbfam=\sevenmsb \scriptscriptfont\msbfam=\fivemsb
        \def\Bbb{\relax\expandafter\Bbb@}
        \def\Bbb@#1{{\Bbb@@{#1}}}
        \def\Bbb@@#1{\fam\msbfam\relax#1}
        \catcode`\@=\active
	
	\font\eightbbb=msbm8
\fi
        \font\eightrm=cmr8              \def\xrm{\eightrm}
        \font\eightbf=cmbx8             \def\xbf{\eightbf}
        \font\eightit=cmti10 at 8pt     \def\xit{\eightit}
        \font\eighttt=cmtt8             \def\xtt{\eighttt}
        \font\eightcp=cmcsc8
        \font\eighti=cmmi8              \def\xold{\eighti}
        \font\teni=cmmi10               \def\old{\teni}
        \font\tencp=cmcsc10
        \font\tentt=cmtt10
        \font\twelverm=cmr12
        \font\twelvecp=cmcsc10 scaled\magstep1
        \font\fourteencp=cmcsc10 scaled\magstep2

\def\noblackbox{\overfullrule=0pt}
\noblackbox

\newtoks\headtext
\headline={\ifnum\pageno=1\hfill\else
{\eightcp\the\headtext}
                \dotfill{ }{\old\folio}\fi}
\def\makeheadline{\vbox to 0pt{\vss\noindent\the\headline\break
\hbox to\hsize{\hfill}}
        \vskip2\baselineskip}
\newcount\infootnote
\infootnote=0
\def\makefootline{\infootnote=1
        \ifnum\foottest>0
                \ifnum\foottest=1
                       \footline={\footlineone}
                \fi
                \ifnum\foottest=2
                        \footline={\footlineone\footlinetwo}
                \fi
                \ifnum\foottest=3
                        \footline={\footlineone\footlinetwo\footlinethree}
                \fi
                \baselineskip=.8cm\vtop{\the\footline}
                \global\foottest=0
        \fi
        \infootnote=0}
\newcount\foottest
\foottest=0
\newdimen\footHsize
\footHsize=\hsize
\advance\footHsize by -\parindent
\advance\footHsize by -79pt
\def\footnote#1#2{${}^#1$\hskip-3pt
	\ifnum\foottest=2
        \def\footlinethree{\hfill\break
        \vtop{\baselineskip=9pt
        \indent ${}^#1$ \vtop{\hsize=\footHsize\noindent\xrm #2}}}\foottest=3
        \fi
        \ifnum\foottest=1
        \def\footlinetwo{\hfill\break
        \vtop{\baselineskip=9pt
        \indent ${}^#1$ \vtop{\hsize=\footHsize\noindent\xrm #2}}}\foottest=2
        \fi
        \ifnum\foottest=0
        \def\footlineone{\vtop{\baselineskip=9pt
        \hrule width.6\hsize\hfill\break
        \indent ${}^#1$ \vtop{\hsize=\footHsize\noindent\xrm #2}}
        \vskip-.7\baselineskip}
        \foottest=1
        \fi
        }
\newcount\refcount
\refcount=1
\newwrite\refwrite
\def\oldsize{\ifnum\infootnote=1\xold\else\old\fi}
\def\ref#1#2{
	\def#1{{{\oldsize\the\refcount}}\ifnum\the\refcount=1\immediate\openout\refwrite=\jobname.refs\fi\immediate\write\refwrite{\item{[{\xold\the\refcount}]} 
	#2\hfill\par\vskip-2pt}\xdef#1{{\oldsize\the\refcount}}\global\advance\refcount by 1}
	}
\def\refout{\catcode`\@=11
        \xrm\immediate\closeout\refwrite
        \vskip2\baselineskip
        {\noindent\twelvecp References}\hfill\vskip\baselineskip
        \baselineskip=.75\baselineskip
        \input\jobname.refs
        \baselineskip=4\baselineskip \divide\baselineskip by 3
        \catcode`\@=\active\rm}

\def\hepth#1{\href{http://xxx.lanl.gov/abs/hep-th/#1}{{\xtt hep-th/#1}}}
\def\jhep#1#2#3{\href{http://jhep.sissa.it/stdsearch?paper=#1\%28#2\%29#3}{{\xit JHEP} {\xbf #1} ({\xold#2}) {\xold#3}}}
\def\PLB#1#2#3{Phys. Lett. {\xbf B#1} ({\xold#2}) {\xold#3}}
\def\NPB#1#2#3{Nucl. Phys. {\xbf B#1} ({\xold#2}) {\xold#3}}
\def\PRD#1#2#3{Phys. Rev. {\xbf D#1} ({\xold#2}) {\xold#3}}
\def\PRL#1#2#3{Phys. Rev. Lett. {\xbf #1} ({\xold#2}) {\xold#3}}
\def\AP#1#2#3{Ann. Phys. {\xbf #1} ({\xold#2}) {\xold#3}}
\def\JGP#1#2#3{J. Geom. Phys. {\xbf #1} ({\xold#2}) {\xold#3}}
\newcount\sectioncount
\sectioncount=0
\def\section#1#2{\global\eqcount=0
	\global\subsectioncount=0
        \global\advance\sectioncount by 1
        \vskip2\baselineskip\noindent
        \line{\twelvecp\the\sectioncount. #2\hfill}
	\vskip\baselineskip\noindent
        \xdef#1{{\old\the\sectioncount}}}
\newcount\subsectioncount
\def\subsection#1#2{\global\advance\subsectioncount by 1
	\vskip\baselineskip\noindent
	\line{\tencp\the\sectioncount.\the\subsectioncount. #2\hfill}
	\vskip.5\baselineskip\noindent
	\xdef#1{{\old\the\sectioncount}.{\old\the\subsectioncount}}}
\newcount\appendixcount
\appendixcount=0
\def\appendix#1{\global\eqcount=0
        \global\advance\appendixcount by 1
        \vskip2\baselineskip\noindent
        \ifnum\the\appendixcount=1
        \hbox{\twelvecp Appendix A: #1\hfill}\vskip\baselineskip\noindent\fi
    \ifnum\the\appendixcount=2
        \hbox{\twelvecp Appendix B: #1\hfill}\vskip\baselineskip\noindent\fi
    \ifnum\the\appendixcount=3
        \hbox{\twelvecp Appendix C: #1\hfill}\vskip\baselineskip\noindent\fi}
\def\acknowledgements{\vskip2\baselineskip\noindent
        \underbar{\it Acknowledgements:}\ }
\newcount\eqcount
\eqcount=0
\def\Eqn#1{\global\advance\eqcount by 1
        \xdef#1{{\oldsize\the\sectioncount}.{\oldsize\the\eqcount}}
        \ifnum\the\appendixcount=0
                \eqno({\oldstyle\the\sectioncount}.{\oldstyle\the\eqcount})\fi
        \ifnum\the\appendixcount=1
                \eqno({\oldstyle A}.{\oldstyle\the\eqcount})\fi
        \ifnum\the\appendixcount=2
                \eqno({\oldstyle B}.{\oldstyle\the\eqcount})\fi
        \ifnum\the\appendixcount=3
                \eqno({\oldstyle C}.{\oldstyle\the\eqcount})\fi}
\def\eqn{\global\advance\eqcount by 1
        \ifnum\the\appendixcount=0
                \eqno({\oldstyle\the\sectioncount}.{\oldstyle\the\eqcount})\fi
        \ifnum\the\appendixcount=1
                \eqno({\oldstyle A}.{\oldstyle\the\eqcount})\fi
        \ifnum\the\appendixcount=2
                \eqno({\oldstyle B}.{\oldstyle\the\eqcount})\fi
        \ifnum\the\appendixcount=3
                \eqno({\oldstyle C}.{\oldstyle\the\eqcount})\fi}
\def\multi{\global\advance\eqcount by 1}
\def\multieq#1#2{\xdef#1{{\oldsize\the\eqcount#2}}
        \eqno{({\oldstyle\the\eqcount#2})}}
\newtoks\url
\def\Href#1#2{\catcode`\#=12\url={#1}\catcode`\#=\active#2}
\def\href#1#2{{#2}}

\parskip=3.5pt plus .3pt minus .3pt
\baselineskip=14pt plus .1pt minus .05pt
\lineskip=.5pt plus .05pt minus .05pt
\lineskiplimit=.5pt
\abovedisplayskip=18pt plus 4pt minus 2pt
\belowdisplayskip=\abovedisplayskip
\hsize=14cm
\vsize=19cm
\hoffset=1.5cm
\voffset=1.8cm
\frenchspacing
\catcode`\@11\relax
\newif\ify@autoscale \y@autoscaletrue \def\Yautoscale#1{\ifnum #1=0
  \y@autoscalefalse\else\y@autoscaletrue\fi}
\newdimen\y@b@xdim
\newdimen\y@boxdim \y@boxdim=13pt
\def\Yboxdim#1{\y@autoscalefalse\y@boxdim=#1}
\newdimen\y@linethick    \y@linethick=.3pt
\def\Ylinethick#1{\y@linethick=#1}
\newskip\y@interspace \y@interspace=0ex plus 0.3ex
\def\Yinterspace#1{\y@interspace=#1}
\newif\ify@vcenter   \y@vcenterfalse
\def\Yvcentermath#1{\ifnum #1=0 \y@vcenterfalse\else\y@vcentertrue\fi}
\newif\ify@stdtext   \y@stdtextfalse
\def\Ystdtext#1{\ifnum #1=0 \y@stdtextfalse\else\y@stdtexttrue\fi}
\newif\ify@enable@skew   \y@enable@skewfalse
\expandafter\ifx\csname enableskew\endcsname\relax
 \y@enable@skewfalse \else \y@enable@skewtrue\fi
\def\y@vr{\vrule height0.8\y@b@xdim width\y@linethick depth 0.2\y@b@xdim}
\def\y@emptybox{\y@vr\hbox to \y@b@xdim{\hfil}}
\ify@enable@skew
 \def\y@abcbox#1{\if :#1\else
   \y@vr\hbox to \y@b@xdim{\hfil#1\hfil}\fi}
 \def\y@mathabcbox#1{\if :#1\else
   \y@vr\hbox to \y@b@xdim{\hfil$#1$\hfil}\fi}
\else
 \def\y@abcbox#1{\y@vr\hbox to \y@b@xdim{\hfil#1\hfil}}
 \def\y@mathabcbox#1{\y@vr\hbox to \y@b@xdim{\hfil$#1$\hfil}}
\fi
\def\y@setdim{%
  \ify@autoscale%
   \ifvoid1\else\typeout{Package youngtab: box1 not free! Expect an
     error!}\fi%
   \setbox1=\hbox{A}\y@b@xdim=1.6\ht1 \setbox1=\hbox{}\box1%
  \else\y@b@xdim=\y@boxdim \advance\y@b@xdim by -2\y@linethick
  \fi}
\newcount\y@counter
\newif\ify@islastarg
\def\y@lastargtest#1,#2 {\if\space #2 \y@islastargtrue
  \else\y@islastargfalse\fi}
\def\y@emptyboxes#1{\y@counter=#1\loop\ifnum\y@counter>0
  \advance\y@counter by -1 \y@emptybox\repeat}
\def\y@nelineemptyboxes#1{%
  \vbox{%
    \hrule height\y@linethick%
    \hbox{\y@emptyboxes{#1}\y@vr}
    \hrule height\y@linethick}\vskip-\y@linethick}
\def\yng(#1){%
  \y@setdim%
  \hskip\y@interspace%
  \ifmmode\ify@vcenter\vcenter\fi\fi{%
  \y@lastargtest#1,
  \vbox{\offinterlineskip
    \ify@islastarg
     \y@nelineemptyboxes{#1}
    \else
     \y@ungempty(#1)
    \fi}}\hskip\y@interspace}
\def\y@ungempty(#1,#2){%
  \y@nelineemptyboxes{#1}
  \y@lastargtest#2,
  \ify@islastarg
   \y@nelineemptyboxes{#2}
  \else
   \y@ungempty(#2)
  \fi}
\def\y@nelettertest#1#2. {\if\space #2 \y@islastargtrue
  \else\y@islastargfalse\fi}
\def\y@abcboxes#1#2.{%
  \ify@stdtext\y@abcbox#1\else\y@mathabcbox#1\fi%
  \y@nelettertest #2.
  \ify@islastarg\unskip%
   \ify@stdtext\y@abcbox{#2}\else\y@mathabcbox{#2}\fi%
  \else\y@abcboxes#2.\fi}
 \newdimen\y@full@b@xdim
 \newcount\y@m@veright@cnt
\ify@enable@skew
 \def\y@get@m@veright@cnt#1#2.{%
   \if :#1 \advance\y@m@veright@cnt by 1\y@get@m@veright@cnt#2.\fi}
 \let\y@setdim@=\y@setdim
 \def\y@setdim{%
   \y@setdim@ \y@full@b@xdim=\y@b@xdim
   \advance\y@full@b@xdim by 1\y@linethick}
 \def\y@m@veright@ifskew#1{
   \y@m@veright@cnt=0 \y@get@m@veright@cnt#1.
   \moveright \y@m@veright@cnt\y@full@b@xdim}
\else
 \def\y@m@veright@ifskew#1{}
\fi
\def\y@nelineabcboxes#1{%
  \y@nelettertest #1.
  \ify@islastarg
   \y@m@veright@ifskew{#1}
    \vbox{
      \hrule height\y@linethick%
      \hbox{\ify@stdtext\y@abcbox#1\else\y@mathabcbox#1\fi\y@vr}
      \hrule height\y@linethick}\vskip-\y@linethick
  \else
   \y@m@veright@ifskew{#1}
    \vbox{
      \hrule height\y@linethick%
      \hbox{\y@abcboxes #1.\y@vr}%
      \hrule height\y@linethick}\vskip-\y@linethick
  \fi}
\def\young(#1){%
  \y@setdim%
  \hskip\y@interspace%
  \y@lastargtest#1,
  \ifmmode\ify@vcenter\vcenter\fi\fi{%
  \vbox{\offinterlineskip
    \ify@islastarg\y@nelineabcboxes{#1}%
    \else\y@ungabc(#1)%
    \fi}}\hskip\y@interspace}
\def\y@ungabc(#1,#2){%
  \y@nelineabcboxes{#1}%
  \y@lastargtest#2,
  \ify@islastarg\y@nelineabcboxes{#2}%
  \else\y@ungabc(#2)%
  \fi}
\catcode`\@12\relax
\Yboxdim6pt

\def\/{\over}
\def\*{\partial}
\def\punkt{\,\,.}
\def\komma{\,\,,}

\def\+{\!+\!}
\def\={\!=\!}
\def\small#1{{\hbox{$#1$}}}
\def\half{\small{1\/2}}
\def\fraction#1{\small{1\/#1}}

\def\eg{{\tenit e.g.}}
\def\ie{{\tenit i.e.}}

\def\w{\!\wedge\!}
\def\id{1\hskip-3.5pt 1}

\def\nl{\hfill\break\indent}
\def\nlni{\hfill\break}

\def\\{\cr}
\def\/{\over}
\def\*{\partial}
\def\a{\alpha}
\def\b{\beta}

\def\d{\delta}
\def\e{\varepsilon}

\def\g{\gamma}
\def\k{\kappa}
\def\l{\lambda}
\def\m{\mu}
\def\n{\nu}
\def\r{\rho}
\def\s{\sigma}

\def\th{\theta}
\def\w{\omega}
\def\W{\Omega}

\def\D{\Delta}
\def\G{\Gamma}
\def\L{\Lambda}
\def\sL{{\cal L}}
\def\M{{\cal M}}
\def\dM{{\*\!\M}}
\def\S{\Sigma}
\def\Z{{\Bbb Z}}
\def\R{{\Bbb R}}
\def\punkt{\,\,.}
\def\komma{\,\,,}

\def\II{\hbox{I\hskip-0.6pt I}}
\def\id{1\hskip-3.4pt 1}
\def\wdg{\!\wedge\!}
\def\fraction#1{\hbox{${1\/#1}$}}
\def\ifrac#1{\hbox{${i\/#1}$}}
\def\Fraction#1#2{\hbox{${#1\/#2}$}}

\def\eg{{\tenit e.g.}}
\def\ie{{\tenit i.e.}}

\def\square{\,\yng(1)\,}

\def\eq#1{$$#1\eqn$$}
\def\eql#1#2{$$#1\Eqn#2$$}

\def\eqal#1#2{$$\eqalign{#1}\Eqn#2$$}

\headtext={Adawi, Cederwall, Gran, Nilsson, Razaznejad: ``Goldstone 
Tensor Modes''}

%
%
%

\null\vskip-1cm
\hbox to\hsize{\hfill G\"oteborg-ITP-98-18}
\hbox to\hsize{\hfill\tt hep-th/9811145}
\hbox to\hsize{\hfill November, 1998}

\vskip3cm
\centerline{\fourteencp Goldstone Tensor Modes}
\vskip4pt
\vskip\parskip
\centerline{\twelvecp}

\vskip1.2cm
\centerline{\twelverm Tom Adawi, Martin Cederwall, Ulf Gran,}
\centerline{\twelverm Bengt E.W. Nilsson and Behrooz Razaznejad}

\vskip.8cm
\centerline{\it Institute for Theoretical Physics}
\centerline{\it G\"oteborg University and Chalmers University of Technology }
\centerline{\it S-412 96 G\"oteborg, Sweden}

\vskip.8cm
\catcode`\@=11
\centerline{\tentt 
	adawi,martin.cederwall,gran,bengt.nilsson,behrooz@fy.chalmers.se}
\catcode`\@=\active

\vskip2.2cm

\centerline{\bf Abstract}

{\narrower\noindent In the context of brane solutions of supergravity, 
we discuss a general method to introduce collective
modes of any spin by exploiting a particular way of breaking
symmetries. The method
is applied to the D3, M2 and M5 branes and we derive 
explicit expressions for how the zero-modes enter the target space
fields, verify normalisability in the transverse directions 
and derive the corresponding
field equations on the brane. In particular,
the method provides a clear understanding of scalar, spinor, and
rank $r$ tensorial Goldstone modes, chiral as well as non-chiral, 
and how they arise
from the gravity, Rarita-Schwinger, and rank $r+1$ Kalb-Ramond tensor
gauge fields, 
respectively. Some additional observations concerning
the chiral tensor modes on the M5 brane are discussed.
\smallskip}
\vfill\eject

\ref\DS{M.J. Duff and K.S. Stelle,
	{\xit ``Multimembrane solutions of D=11 supergravity''},
	\nl\PLB{253}{1991}{113}.}
\ref\Kaplan{D.M. Kaplan and J. Michelson, {\xit ``Zero modes for the D=11 
	membrane and five-brane''}, 
	\nl\PRD{53}{1996}{3474} [\hepth{9510053}].}
\ref\Bagger{J. Bagger and A. Galperin,
	{\xit ``New Goldstone multiplet for partially broken supersymmetry''},
	\nl\PRD{55}{1997}{1091} [\hepth{9608177}];\nlni
	{\xit ``The tensor Goldstone multiplet for partially 
		broken supersymmetry''},
	\nl\PLB{412}{1997}{296} [\hepth{9707061}].}
\ref\NextPaper{M. Cederwall, U. Gran, M. Holm and B.E.W. Nilsson,
	in preparation.}
\ref\KappaBranes{M.~Cederwall, A.~von~Gussich, B.E.W.~Nilsson 
	and A.~Westerberg,\nl
	{\xit ``The Dirichlet super-three-brane in type \II B supergravity''},
	\nl \NPB{490}{1997}{163} [\hepth{9610148}];\nlni
	M. Cederwall, A. von Gussich, B.E.W. Nilsson, P. Sundell
        and A. Westerberg, \nl{\xit ``The Dirichlet super-p-branes in
        type \II A and \II B supergravity''}, 
	\nl\NPB{490}{1997}{179} [\hepth{9611159}];\nlni
	M. Aganagic, C. Popescu and J.H. Schwarz, 
	{\xit ``D-brane actions with local kappa symmetry''}, 
	\nl\PLB{393}{1997}{311} [\hepth{9610249}];\nl
        {\xit ``Gauge-invariant and gauge-fixed D-brane actions''},
        \NPB{495}{1997}{99} [\hepth{9612080}];\nlni
	E. Bergshoeff and P.K. Townsend, {\xit ``Super D-branes''}, 
	\NPB{490}{1997}{145} [\hepth{9611173}].}
\ref\CT{M. Cederwall and P.K. Townsend, {\xit ``The manifestly 
	{\xrm SL(2;{\eightbbb Z})}-covariant superstring''}, 
	\nl\jhep{09}{1997}{003} [\hepth{9709002}].}
\ref\CW{M. Cederwall and A. Westerberg,
	{\xit ``World-volume fields, {\xrm SL(2;{\eightbbb Z})} and duality: 
		the type \II B 3-brane''},
	\nl\jhep{02}{1998}{004} [\hepth{9710007}].} 
\ref\CNS{M. Cederwall, B.E.W. Nilsson and P. Sundell, 
	{\xit ``An action for the 5-brane in D=11 supergravity''},
	\nl\jhep{04}{1998}{007} [\hepth{9712059}].} 
\ref\CHS{C.G. Callan, Jr., J.A. Harvey and A. Strominger,
	{\xit ``Worldbrane actions for string solitons''}, 
	\nl\NPB{367}{1991}{60}.}
\ref\ElevenSG{E. Cremmer, B. Julia and  J. Scherk,
	{\xit ``Supergravity theory in eleven-dimensions''},
	\nl\PLB{76}{1978}{409};\nlni
	L.~Brink and P.~Howe, {\xit ``Eleven-dimensional supergravity 
	on the mass-shell in superspace''},
	\nl\PLB{91}{1980}{384};\nlni
	E.~Cremmer and S.~Ferrara, 
	{\xit ``Formulation of eleven-dimensional supergravity 
	in superspace''},\nl\PLB{91}{1980}{61}.}
\ref\Duality{E.~Witten, 
	{\xit ``String theory dynamics in various dimensions''},
	\nl\NPB{443}{1995}{85} [\hepth{9503124}];\nlni
	C.M. Hull and P.K. Townsend,
        {\xit ``Unity of superstring dualities''},
        \nl\NPB{438}{1995}{109} [\hepth{9410167}];\nlni
	J.H. Schwarz, {\xit ``The power of M theory''},
        \PLB{367}{1996}{97} [\hepth{9510086}];\nlni
	A.~Sen, {\xit ``Unification of string dualities''},
	Nucl. Phys. Proc. Suppl. {\xbf58} ({\xold1997}) {\xold5}
	[\hepth{9609176}];\nlni
	P.K. Townsend, {\xit ``Four lectures on M-theory''},
        \hepth{9612121}.}
\ref\Guven{R.~G\"uven, 
	{\xit ``Black p-brane solutions of D=11 supergravity theory''}, 
	\PLB{276}{1992}{49}.}
\ref\Fivebranes{E. Witten, {\xit ``Five-brane effective action in M-theory''},
	\JGP{22}{1997}{103} [\hepth{9610234}];\nlni
	M. Aganagic, J. Park, C. Popescu and J.H. Schwarz,\nl
	{\xit ``World-volume action of the M theory five-brane''},
	\NPB{496}{1997}{191} [\hepth{9701166}];\nlni
	T. Adawi, M. Cederwall, U. Gran, M. Holm and B.E.W. Nilsson,\nl
	{\xit ``Superembeddings, non-linear supersymmetry and 5-branes''},
	\hepth{9711203}, \nl to appear in Int. J. Mod. Phys. {\xbf A};\nlni
	P.S.~Howe and E.~Sezgin, {\xit ``D=11, p=5''},
        \PLB{394}{1997}{62} [\hepth{9611008}];\nlni
        P.S.~Howe, E.~Sezgin and P.C.~West,\nl
        {\xit ``Covariant field equations of the M theory five-brane''},
        \PLB{399}{1997}{49} [\hepth{9702008}];\nl
        {\xit ``The six-dimensional self-dual tensor''},
        \PLB{400}{1997}{255} [\hepth{9702111}];\nlni
	E. Bergshoeff, M. de Roo and T. Ortin,      
        {\xit ``The eleven-dimensional five-brane''},\nl
        \PLB{386}{1996}{85} [\hepth{9606118}].}
\ref\StringSolitons{M.J. Duff, R.R. Khuri and J.X. Lu,
	{\xit ``String solitons''},
	Phys. Report. {\xbf259} ({\xold1995}) {\xold213}  [\hepth{9412184}].}
\ref\HoweWest{P.S. Howe and P.C. West,
	{\xit ``The complete N=2, d=10 supergravity''},
	\NPB{238}{1984}{181}.}
\ref\Rajaraman{R. Rajaraman, {\xit ``Solitons and instantons''},
	North-Holland, 1982;\nlni
	J.A. Harvey, {\xit ``Magnetic monopoles, duality and supersymmetry''}, 
	\hepth{9603086}.}
\ref\ElevenSG{E. Cremmer, B. Julia and  J. Scherk,
	{\xit ``Supergravity theory in eleven-dimensions''},
	\nl\PLB{76}{1978}{409};\nlni
	L.~Brink and P.~Howe, {\xit ``Eleven-dimensional supergravity 
	on the mass-shell in superspace''},
	\nl\PLB{91}{1980}{384};\nlni
	E.~Cremmer and S.~Ferrara, 
	{\xit ``Formulation of eleven-dimensional supergravity 
	in superspace''},\nl\PLB{91}{1980}{61}.}
\ref\DuffLu{M.J. Duff and J.X. Lu, {\xit ``The self-dual type \II B
	superthreebrane''}, \PLB{273}{1991}{409}.}
\ref\DuffReview{M.J. Duff, {\xit ``Supermembranes''}, \hepth{9611203}.}
\ref\FivebraneLagr{
	I. Bandos, K. Lechner, A. Nurmagambetov, P. Pasti, D. Sorokin
        and M. Tonin,\nl {\xit ``Covariant action for the super-five-brane
                of M-theory''},
        \nl\PRL{78}{1997}{4332} [\hepth{9701037}].}
\ref\Supermembrane{E.~Bergshoeff, E.~Sezgin and P.K.~Townsend, 
	\nl {\xit ``Supermembranes and eleven-dimensional supergravity''},
	\PLB{189}{1987}{75};
	\nl {\xit ``Properties of the eleven-dimensional supermembrane 
	Theory''}, \AP{185}{1988}{330}.}
\ref\SorokinTownsend{D. Sorokin and P.K. Townsend,
    {\xit ``M-theory superalgebra from the M-5-brane''},
    \nl\PLB{412}{1997}{265} [\hepth{9708003}].}
\ref\StringSoliton{P.S. Howe, N.D. Lambert and P.C. West,
    {\xit ``The self-dual string soliton''},
    \nl\NPB{515}{1998}{203} [\hepth{9709014}];\nlni
	 J.P. Gauntlett, N.D. Lambert and P.C. West,
	{\xit ``Supersymmetric fivebrane solitons''},
	\hepth{9811024}.}


\section\intro{Introduction}In theories with monopole or instanton solutions
the study of moduli and collective coordinates
has a long and interesting history. Quite generally two different
kinds of moduli appear. {\it E.g.} in the context of the SU(2) 
gauge theory with
a Higgs field one finds moduli describing the freedom to locate the 
monopole anywhere in space as well as one modulus stemming from the
abelian gauge symmetry surviving the symmetry breaking.
While moduli of the former kind are easily introduced
by shifting the space coordinates, generating so called collective coordinates,
the latter kind requires a more detailed analysis
of the gauge theory itself. As described \eg\ in [\Rajaraman], 
the fourth modulus of the
BPS monopole arises from a special choice of gauge parameter corresponding
to a large gauge transformation. In the following we will refer to both 
kinds of zero-modes as collective modes, while their constant part will be 
considered as moduli. Since we will be dealing with extended objects and their
zero-modes, we do not find it fruitful to make a distinction between 
static and non-static configurations. In order to maintain the covariance
of the dynamics of the zero-modes, which of course describe a field theory
``on the brane'', it is more fruitful to generalise the concept of 
motion on moduli space, relevant for point-like solitons, to variation
of the collective coordinates with any of the (timelike or spacelike)
longitudinal directions.

In the recent non-perturbative developments in string theory
(see \eg\ ref. [\Duality]), 
these issues must be reexamined in the context of the 
$p$-, D-, and tensor branes appearing 
as solitonic solutions of various M-theory supergravity theories
(for a review, see \eg\ ref. [\DuffReview]). The tensor
brane M5, in particular, contains as collective modes 
self-dual tensor fields in six dimensions 
[\Fivebranes,\CNS,\FivebraneLagr],
and will thus constitute a slightly more complicated but at the same time much
more interesting
example of these ideas. Common to all branes appearing as solutions to
supergravity (in contrast to solitons in field theory without gravity)
is the feature that all their collective modes are related to broken
gauge symmetries and, as we will explain in detail below for the 
D3, M2 and M5 branes, 
these modes can be extracted from the target space gauge fields by
making a judicious choice of the relevant gauge parameter.

Although the nature of the zero-modes discussed here has been known for
some time and has been used in a number of applications, their 
explicit relationship
to the brane solutions of supergravity has only been briefly touched upon
[\CHS,\Kaplan]. We find it important that this situation is improved, so
that the understanding of these aspects of string theory/M-theory solitons
is put on a more equal footing to that of solitons in field theory.

The paper is organised as follows. 
In the next section we recapitulate the properties of $D=11$ supergravity 
[\ElevenSG] and type IIB supergravity in $D=10$ [\HoweWest] 
that will be needed in subsequent sections.
This section also sets the notation and introduces the various brane solutions
on which we will focus our attention, namely the D3, M2 and M5 branes.
Section {\old3} then describes the procedure which will tell us how
the collective modes emerge from the target space fields. From this 
procedure it will also be clear in what sense the collective modes are related
to broken symmetries. In particular, we will discover how self-dual
gauge fields in six dimensions can arise from the three-form potential
of $D=11$ supergravity. The order in which the collective modes are discussed
is scalar, spinor, vector, and self-dual tensor. Section {\old3} 
ends by some comments on normalisability and other issues. In section 
{\old4}, some specific questions connected to the excitation of 
tensorial zero-modes
are discussed, \eg\ their electric charge. Section {\old5} contains a 
summary and some further comments.

\section\prel{Preliminaries}The purpose of this section is to set 
the stage for the subsequent discussions
of the M2 and M5 branes of $D=11$ supergravity and the D3 brane in type 
IIB $D=10$ supergravity. We will therefore start by reviewing 
these solutions and the field equations they solve. The conventions we use are
listed in the appendix.

The bosonic action of eleven-dimensional supergravity is 
\eql{S=\int d^{11}x\sqrt{-g}\left(R-\fraction{48}H_{MNPQ}H^{MNPQ}\right)
	+\int\fraction{6}H\wdg H\wdg C\komma
	}{\lag} 
where the 4-form $H=dC$, which gives rise to the equations of motion
\eql{	R_{MN}-\fraction{2} g_{MN} R=
	\fraction{12}H_{MPQR}H_N{}^{PQR} 
	-\fraction{96}g_{MN}H_{PQRS}H^{PQRS}}{\stresstensor} 
and
\eql{	D^MH_{MNPQ}={\sqrt{|g|}\/2(4!)^2}
	\e_{NPQ}{}^{R_1\ldots R_8}H_{R_1\ldots R_4}H_{R_5\ldots R_8}\komma
	}{\Heqv}
\smallskip\noindent
or, equivalently, $d{\star}H= \fraction{2} H\wdg H$.
It is often convenient to rewrite the first equation of
motion as
\eql{	R_{MN} = \fraction{12}H_{MPQR}H_N{}^{PQR}
    -\fraction{144}g_{MN}H_{PQRS}H^{PQRS}\punkt
	}{\eqmrw} 

When we look at a specific brane solution we split
the $M$ index into $(\m,m)$, where $\m$ denotes a direction on
the brane and $m$ a direction transverse to the brane.
For the extremal 2-brane [\Supermembrane] the solution [\DS] is
\eqal{	ds^2&=\D^{-{2\/3}}\eta_{\m\n}dx^\m dx^\n
	+\D^{{1\/3}}\d_{pq}dy^pdy^q\komma\\
	C&=\pm\fraction{3!}\D^{-1}\e_{\m\n\r}dx^\m\wdg dx^\n\wdg dx^\r\komma
	}{\twosol} 
where
\eql{	\D=1+\Bigl({R\/\r}\Bigr)^6
	}{\lamb} 
and
\eq{	\r=\sqrt{\d_{mn}y^my^n}\punkt
	} 
  
The corresponding extremal 5-brane solution [\Guven] is
\eqal{	ds^2&=\D^{-{1\/3}}\eta_{\m\n}dx^\m
	dx^\n+\D^{{2\/3}}\d_{mn}dy^mdy^n\komma\\
	H&=\pm\fraction{4!}\d^{mn}\*_m\D\e_{npqrs}dy^p\wdg
	dy^q\wdg dy^r\wdg dy^s\komma }{\fivesol} 
where $\D$ now is defined as
\eql{	\D=1+\Bigl({R\/\r}\Bigr)^3\punkt
	}{\delt} 
Both solutions are given in so called isotropic coordinates, where the
isotropy groups SO(1,$p$) $\times$SO(11-$p$-1) are manifest and $\r=0$ is the
location of the horizon.
The harmonic property of the
$\D$'s, $\d^{pq}\*_p\*_q\D$=0, 
is all that is needed to verify the solutions (\twosol)
and (\fivesol). The sign of the tensor field signifies positive or
negative charge, \ie, a brane or an anti-brane. In the sequel, the
positive sign will be chosen. The other sign will imply a switch of the
chiralities of the zero-modes as will be clear in the following section.
These comments also apply to the D3 extremal solution to which we now turn. 

In the case of type IIB supergravity we give only the field equations
to avoid the at this stage irrelevant discussion of actions for 
self-dual gauge fields. Here, and in the discussion of the zero-modes, 
we will need the field equations for
the metric, the 2-form complex tensor potential $B$ and
the 4-form potential $C$ with self-dual field strength. 
We simplify the calculation by the
initial observation that the scalar fields, taking values in the
coset SL(2;$\R$)/U(1), are constant in the D3 brane
solution; the associated connections then vanish. 
The field strength of $C$ is 
$G=dC+i(\bar B\wdg H-B\wdg\bar H)$, and, given that the scalars are constant,
the complex 3-form field strength is $H=dB$.

The 2-form potential does not enter until we consider deformations of
the D3 brane solution, so the relevant information in Einstein's equation 
reads

\eql{R_{MN}=\fraction{96}G_{MPQRS}{G_N}^{PQRS}+\hbox{other fields}\komma
}{\einIIB}
while the equations for the tensors are

\eql{G_{MNPQR}={\star}G_{MNPQR}}{\IIBeqone}
and
$$
d{\star}H+iG\wdg H=0
$$
(the last of these again makes use of the vanishing of the connections
built from the scalars). 

The D3 brane solution is now given by [\DuffLu]

\eqal{	ds^2&=\D^{-{1\/2}}\eta_{\m\n}dx^\m
	dx^{\n}+\D^{{1\/2}}\d_{mn}dy^mdy^n\komma\\
	G&=\pm\fraction{5!}(\d^{mn}\*_m\D\e_{npqrst}dy^p\wdg
	dy^q\wdg dy^r\wdg dy^s\wdg dy^t\\
	&\quad\thinspace+5\*_m\D^{-1}\e_{\m\n\r\s}dy^m\wdg dx^{\m}\wdg
	dx^{\n}\wdg dx^{\r}\wdg dx^{\s})\\
	}
	{\threesol} 
and $H=0$, where
\eql{\D=1+\Bigl({R\/\r}\Bigr)^4\punkt
	}{\deltDthree} 

From now on all contractions are done with flat metric 
tensors which will not appear explicitly.

\section\zm{Zero-modes}The presence in spacetime of any object, 
like the extended ones discussed in the previous
section, breaks some of the symmetries of the 
background. The breaking
of these symmetries gives rise to Goldstone modes living on the
branes. Since we are dealing with a supersymmetric theory, there will be both
fermionic and bosonic Goldstone modes.
Furthermore, since the branes discussed here leave half of the spacetime
supersymmetry unbroken the Goldstone modes will fall into ordinary
supermultiplets, for which the
number of fermionic and bosonic modes are equal. The broken supersymmetries
of the M2 brane
solution give rise to eight Goldstone
fermions, while the broken translational symmetry in the transverse directions,
leads to eight Goldstone scalars. The M5 brane solution also breaks
half of the supersymmetry, but it has only five transverse
directions. The breaking of the translational symmetry in these
transverse directions gives just five Goldstone scalars and thus there 
are three bosonic zero-modes missing. These bosonic zero-modes 
come from an (anti-)self-dual 3-form and arise from
breaking the gauge symmetry of the background 3-form potential
$C$ in exactly the same way as for the scalar and fermionic modes.

The viewpoint presented in the previous paragraph conforms with the 
standard picture of a BPS brane in flat space, breaking half
of the rigid supersymmetries as well as the transverse translations. From
the supergravity point of view, these are global symmetries of the asymptotic
Minkowski region in the solutions of section \prel. In the supergravity
theory {\it per se}, without assuming a specific background, it is not
meaningful to talk about global symmetries in this sense---all relevant
symmetries are local (reparametrisations, local supersymmetry, tensor
gauge symmetry). The discussion that follows identifies the parts
of these symmetries that are relevant for the Goldstone mechanism in
brane solutions and the properties of the solutions that are essential
for the mechanism.

We discuss a method, which can be used to obtain all the zero-modes
for both the M2 and M5 brane in $D=11$ and the D3 brane in $D=10$. The main
idea is to start from
global, or large, gauge transformations on the background fields. 
The precise sense in which the transformations are large is that the gauge
parameters take different values in the asymptotic Minkowski region,
$\r>\!>R$, and close to the horizon, $\r<\!<R$.
This step thus
introduces the dependence on the moduli for all the fields that are
affected by the transformation. By making
these transformations local in the brane coordinates and requiring 
that the transformed fields
satisfy the equations of motion, we obtain the transversal 
behavior of the target space fields and the
equations of motion for the zero-modes, or collective coordinates as they will
be called here. Now that we have put
forward the general idea, we proceed to do the calculations.

\subsection\ScalarZM{The Scalar Zero-modes}We first consider 
the scalar zero-modes. Since these modes
are related to the breaking of the translational symmetry in the
transverse directions, the relevant symmetries are 
infinitesimal diffeomorphisms. Under
such transformations the metric changes as
\eql{	h_{MN}=\d g_{MN}=\sL_\e g_{MN}=2D_{(M}{\e}_{N)}\komma
	}{\metricvar}
where $M=(\mu,m)$ corresponds to the split into brane and transverse 
directions coordinatised by $x^{\mu}$ and $y^m$, respectively.
We now want to compute the change that results from a 
coordinate transformation transverse to the brane with parameter
$\e^{m}=\D^{s}{\bar \phi}^{m}$ (and $\e^{\mu}=0$), where $s$ is a parameter 
to be determined and 
${\bar{\phi}}^{m}$ are {\it constant} moduli. 
In order to have an expression useful for all the cases
under discussion we give the answer for 
general values of the dimensions $D$ of the target space and $d=p+1$
of the brane, and general parameters $\alpha$ and $\b$ in the metric Ansatz

\eql{	ds^2=\D^{2\a}dx^{2}
+\D^{2\b}dy^2\punkt}{\genmetr}
This gives

\eqal{	h^{(mod)}_{\mu\nu}&=
2\alpha \D^{s+2\alpha-1}({\bar {\phi}}^p{\partial}_p{\D})
{\eta}_{\mu\nu}\komma\\
h^{(mod)}_{\mu n}&=0\komma\\
h^{(mod)}_{mn}&=
2\D^{s+2\b-1}(s{\bar {\phi}}_{(m}{\partial}_{n)}{\D}+
\b\d_{mn}{\bar {\phi}}^p{\partial}_p{\D})\komma
}{\genh}
where the superscript $(mod)$ indicates that the gauge parameters
$\bar{\phi}$ are constant.
Note that if ${\bar \phi}^m$ had been functions 
on the brane at this stage,
there would have been additional derivative terms in these expressions. 

We now drop the bar on ${\bar \phi}^m$ 
and let them become functions on the brane: $\phi^m=\phi^m(x)$.
These functions are the zero-modes and will from now on be
referred to as collective coordinates $(cc)$. The 
corresponding metric components are denoted $h^{(cc)}_{MN}$. They are
still given by eq. (\genh), and {\it they are no longer
pure gauge transformations}. This last fact is the reason why
new  physical modes appear in the theory. That they are exactly the zero-modes
we are interested in will now be shown.

Having obtained the proper form of the metric Ansatz, we should now insert it
into the Einstein equations. Since we will work only to linear order
in the perturbations away from the brane solution, the following observation
will be calculationally useful. Since both a given background 
and the background changed by a gauge transformation with
constant ${\bar \phi}^m$
solve the field equations, only terms containing at least one $x$-derivative
on $\phi^m(x)$ will survive. The variation of the Ricci tensor reads
\eql{	
\d{R}^{(cc)}_{MN}=-\Fraction{1}{2}{\nabla^Q\nabla_Q}h^{(cc)}_{MN}+
\nabla_{(M}\nabla^Q h^{(cc)}_{N)Q}-
\Fraction{1}{2}\nabla_{(M}\nabla_{N)}h^{(cc)Q}_Q+\;
\hbox{non-derivative terms}\punkt
}{\RicVarD}
Inserting the above metric Ansatz then gives 
\eqal{	
\d{R}^{(cc)}_{\mu\nu}&=-\alpha{\D}^{s-1}{\eta}_{\mu\nu}
	({\*}^{\r}{\*}_{\r}\phi^{m}){\*}_{m}{\D}\\
&-[s+\a (d-2)+\b(D-d)]
	{\D}^{s-1}({\*}_{\mu}{\*}_{\nu}\phi^{m})\*_m{\D}
	\komma\\
\d{R}^{(cc)}_{mn}&=-{\D}^{s+2\b-2\a-1}
		[s({\*}^{\mu}{\*}_{\mu}\phi_{(m}){\*}_{n)}{\D}
	+\b{\d}_{mn}({\*}^{\mu}{\*}_{\mu}\phi^{p}){\*}_{p}{\D}]\komma\\
\d{R}^{(cc)}_{\mu n}&=\Fraction{s}{2}{\D}^{s-1}\*_{\m}\phi_n \*^m \*_m \D\\ 
&+\bigl(\Fraction{s}{2}-\a+\b-[s+\a (d-2)+\b(D-d)]\bigr){\D}^{s-1}
	{\*}_{\mu}\phi^p{\*}_{n}{\*}_p{\D}\\
&+\half s[s +\a (d-2)+\b(D-d)-1]{\D}^{s-2}
	{\*}_{\mu}{\phi_n}{\*}^{m}{\D}{\*}_m{\D}\\
&+\bigl(\a-\b+\a\b(D-2)-\Fraction{s}{2}
	+(1-\Fraction{s}{2})[s+\a (d-2)+\b(D-d)]\bigr)\\
&\hskip1cm\times{\D}^{s-2}{\*}_{\mu}{\phi_m}{\*}^{m}{\D}{\*}_n{\D}\punkt
	}{\Ricvarcomp}

The variation of the Ricci tensor should now be equated to the variation
of the RHS of (\eqmrw). In fact, we get no
contribution (with longitudinal derivatives) 
from the second term in the RHS of (\eqmrw), which can be seen by
considering the index structure\footnote{\dagger}{Therefore, it is not
essential to distinguish the RHS's of eqs. (\stresstensor) and (\eqmrw), 
and the latter will also be referred to as the stress tensor.}. When computing 
$\d T^{(cc)}_{MN}$ one immediately
realises that $\phi^m$ will never appear acted on by two $x$-derivatives,
and also, by considering the index structure, that both $\d T^{(cc)}_{\m\n}$ 
and $\d T^{(cc)}_{mn}$ are zero
modulo $\phi^m$ terms without $x$-derivatives. Using (\Ricvarcomp), 
$\d R^{(cc)}_{\m\n}=0$
then implies
\eq{\square\phi\equiv\*^{\m}\*_{\m}\phi^n=0}
and
\eql{	s+\a (d-2)+\b (D-d)=0\punkt}{\gencond}
The first of these conditions also means that $\d R^{(cc)}_{mn}=0$ is 
satisfied. It is interesting to note that for the branes we consider
the parameters related to the metric Ansatz satisfy

\eql{	\a (d-2)+ \b (D-d)=1\komma}{\metriccond}
implying that $s=-1$ in all cases. Eq. (\metriccond) is in fact a well-known 
condition which guarantees that no velocity dependent forces appear between
branes
when the brane action is expanded to lowest non-trivial order in the
collective modes related to broken translations [\StringSolitons]. 

Before turning to the actual computation
of the stress tensor variation we plug the condition (\gencond) 
into the 
the variation of the Ricci tensor. This produces the much simpler expressions:
\eqal{	
\d{R}^{(cc)}_{\mu\nu}&=-\alpha{\D}^{s-1}{\eta}_{\mu\nu}
\square\phi^{m}{\*}_{m}{\D}
	\komma\\
\d{R}^{(cc)}_{mn}&=-{\D}^{s+2\b-2\a-1}
		(s\square\phi_{(m}{\*}_{n)}{\D}
	+\b{\d}_{mn}\square\phi^{p}{\*}_{p}{\D})\komma\\
\d{R}^{(cc)}_{\mu n}&={\D}^{s-1}
(\hbox{${s\/2}$}\*^{\m}\phi_n \*^\m \*_m \D
(\hbox{${s\/2}$}-\a+\b){\*}_{\mu}\phi^p)
{\*}_{n}{\*}_p{\D})\\
&-\half s{\D}^{s-2}
	{\*}_{\mu}{\phi_n}{\*}^{m}{\D}{\*}_m{\D}\\
&+(\a-\b+\a\b(D-2)-\hbox{${s\/2}$}){\D}^{s-2}
	{\*}_{\mu}{\phi_m}{\*}^{m}{\D}{\*}_n{\D}\punkt}{\Ricvarcompsimp}

To verify the last of  Einstein's equations we now derive the expression for
the linearised stress tensor for the three different cases under discussion.
We start by considering $D=11$ supergravity and its stress tensor given in 
equation (\stresstensor). Both the metric and the 3-form potential 
should now be varied
under coordinate transformations. We will however not get any relevant
contribution from the variation of the metric since there is no derivative
acting on it. The variation of $C_{MNP}$ follows from
\eql{\d_{\e}C=\sL_\e C+d\L=(i_{\e}d+di_{\e})C+d\L=i_{\e}H}{\tensorvar}
obtained by choosing the accompanying 
gauge transformation 2-form parameter $\L=-i_{\e}B$. 
Here $H$ is the background value which 
means that for the M5 brane solution the only non-zero components of $\d C$ are
\eq{\d_{\e}C_{mnp}=-\D^s\phi^q \*^r\D\e_{qmnpr}\punkt}
As for the Ricci tensor only $\*_{\m}\phi^m$ terms need be kept in 
$\d T^{(cc)}_{MN}$ when inserted into the field equations. Hence, for the M5
brane we get
\eqal{	\d T^{(cc)}_{\m\n}&=\d T^{(cc)}_{mn}=0\komma\\
	\d T^{(cc)}_{\m n}&=\half{\D}^{s-2}(\*_{\m}\phi_n\*^p{\D}\*_p{\D}
-\*_{\m}\phi^m\*_m{\D}\*_n{\D})}{\MfiveT}
and we actually get the same result for the M2 and D3 brane.
In order to do the calculation for the M2 brane we use the dual formulation,
where the stress tensor is given by
\eq{T_{MN}=\fraction{6\cdot6!}H_{M P_1 \ldots P_6}H_{N}{}^{P_1 \ldots
P_6}-\fraction{12\cdot7!}g_{MN}H_{P_1 \ldots P_7}H^{P_1 \ldots P_7}
	}
and the non-vanishing component of the variation is
\eq{	\d_{\e}H_{\m p_1 \ldots p_6}=\*_{\m}\phi^q\e_{q p_1 \ldots p_6 r}
\D^{s}\*^r \D
	\komma}
which we obtain as in (\tensorvar).

Finally, in order to do the calculation for the D3 brane, we use the 
expression for the type IIB, $D=10$ stress tensor given in
equation (\einIIB) and that the variation, obtained as before, is
\eq{\d_{\e}G_{\m p_1 \ldots p_4}=\*_\m \phi^m \*^n 
\D \e_{n m p_1 \ldots p_4} \D^s
	}
and
\eq{\d_{\e}G_{\m_1 \ldots \m_5}=-5\*_{[\m_1}\phi^m \*_m \D^{-1} 
\e_{\m_2 \ldots \m_5]}\D^s
	\punkt}
However, the self-duality of $G$ requires that we also have
\eq{\d_{\e}G_{m n \m \n \r}=-2\*^{\s}\phi_{[m}\*_{n]}
\D \e_{\s \m \n \r}\D^{s-1}
	\punkt}

Using the various values for the parameters $D,d,\a,\b$ for the 
three cases under study one concludes that
the Einstein equations are all satisfied provided $\square\phi^m=0$, 
$s=-1$ and that the function $\D$ is
harmonic in the transverse coordinates, \ie, that $\*^m\*_m{\D}=0$.

In order to check the normalisability of the bosonic zero-modes we 
integrate out
the transversal dependence of the $R$ term in the action, thus obtaining an
effective world-volume action for the zero-modes. We find that for all
three branes the zero-modes are normalisable. 
From now on, the superscripts
$(mod)$ and $(cc)$ will be suppressed.

\vfill\eject
\subsection\FermZM{The Fermionic Zero-modes}We now turn to the 
Goldstone fermions. The supersymmetry transformation
in $D=11$ supergravity 
is given by\footnote*{Since the background is purely bosonic, we drop
all higher order terms in the gravitino field in transformations, covariant
derivatives and equations of motion.}
\eql{ 	\d{\psi}_{M}={\tilde{D}}_{M}{\zeta}={D}_{M}{\zeta}
	-\fraction{288}({\G}_{M}{}^{NPQR}
	-8\d_M{}^N{\G}^{PQR}){\zeta}H_{NPQR}\komma
	}{\superDeleven}
and in type IIB, $D=10$ supergravity, for vanishing $B$-field, by
\eql{	 \d{\psi}_{M}={\tilde{D}}_{M}{\zeta}={D}_{M}{\zeta}
	-\ifrac{192}{\G}^{NPQR}{\zeta}G_{MNPQR}\komma}{\superDthree}
which for convenience we write as
\eql{	\d{\psi}={\tilde{D}}{\zeta}={\*}{\zeta}+{\w}{\zeta}+{\chi}{\zeta}\komma
	}{\supershort} 
the three terms denoting the derivative
term, the spin connection term and
the $H$- or $G$-term, respectively. The general expressions for the 
spin connections are
\eqal{	\omega_{\m}&={1\/4}\omega_{\m AB}{\G}^{AB}=
\half\a {\D}^{\a-\b-1}\G_{\m}\G^m\*_m{\D}\komma\\
	\omega_{m}&={1\/4}\omega_{mAB}{\G}^{AB}=
\half\b {\D}^{-1}\G_m{}^n\*_n{\D}}{\spinconn}
and we split the $\G$ matrices according to
\eqal{	\G_{A}=(\g_{\a}\otimes{\S^9},\id\otimes\S_{a})
	\komma\\
	\G_{A}=(\g_{\a}\otimes{\id},\g^{7}\otimes\S_{a})
	\komma\\
	\G_{A}=(\g_{\a}\otimes{\id},\g^{5}\otimes\S_{a})
	\komma}{\gammasplitgen}
for the M2, M5 and D3 brane, respectively. This split corresponds to splitting
	the group SO(1,$D-1$) into SO(1,$d-1$)$\times$SO($D-d$), \ie, into
	longitudinal and transverse directions.

We now start with the M5 brane and also take into account the $\chi$
terms. Using the M5 brane solution
(\fivesol) and the split of the $\G$ matrices, we obtain the following 
expressions for the spin
connection terms and the $H$-terms given in equation (\supershort):
\eqal{	
{\w}_{\mu}{\zeta}&=
	-\Fraction{\th}{12}{\D}^{-{3\/2}}
	{\*}_{m}{\D}{\g}_{\bar{\mu}}\S^{\bar{m}}{\zeta}\komma\\
{\w}_{m}{\zeta}&=\fraction{6}{\D}^{-1}{\*}_{n}{\D}{\S}_{\bar{m}}
	{}^{\bar{n}}{\zeta}\komma\\
{\chi}_{\mu}{\zeta}&=-\fraction{12}{\D}^{-{3\/2}}
	{\*}_{m}{\D}{\g}_{\bar{\mu}}{\S}^{\bar{m}}{\zeta}\komma\\
{\chi}_{m}{\zeta}&=-\Fraction{\th}{12}{\D}^{-1}{\*}_{m}{\D}{\zeta}
	+\Fraction{\th}{6}{\D}^{-1}{\*}_{n}
	{\D}{\S}_{\bar{m}}{}^{\bar{n}}{\zeta}\komma
	}{\spinterms} 
where $\th=\pm{1}$ is the 6-dimensional chirality of $\zeta$ ($\g_7$ has 
simply been replaced by its eigenvalue $\th$).
Overlined indices are inertial indices---we prefer to use inertial
$\G$-matrices in order to manifest explicitly all radial dependence. 
A gauge transformation
with $\zeta={\D}^{k}{\l}$, where $\l$ is a
constant spinor, gives
\eqal{
{\d}{\psi}_{\mu}&=-\Fraction{\th+1}{12}{\D}^{k-{3\/2}}
	{\S}^{\bar{m}}{\*}_{m}{\D}{\g}_{\bar{\mu}}{\l}\komma\\
{\d}{\psi}_{m}&=(k-\Fraction{\th}{12}){\D}^{k-1}{\*}_{m}{\D}{\l}
	+\Fraction{\th+1}{6}{\D}^{k-1}{\*}_{n}
	{\D}{\S}_{\bar{m}}{}^{\bar{n}}{\l}\punkt
	}{\fivepot} 
From the equations above we see that the surviving
supersymmetry, obeying the Killing spinor equation $\d\psi=0$, 
has $\th=-1$ and $k=-{1\/12}$. To obtain
the Goldstone fermions we let the gauge parameter $\l$ 
be $x$-dependent and require $\psi$, still given by eq. (\fivepot), 
to satisfy the equation of motion
$$
T_M\equiv\G^NT_{MN}=0\komma\Eqn\PsiEOM
$$
where $T_{MN}=2\tilde{D}_{[M}{\psi}_{N]}$ is the field strength of $\psi$. 
Eq. (\PsiEOM) is equivalent to the usual Rarita--Schwinger equation
$\G^{MNP}\tilde{D}_N\psi_P=0$, but easier to handle.
Since by performing a global gauge
transformation on a solution of the equations of motion we just obtain
another solution, only the ${\*}_{\mu}{\l}$-terms in the field 
strength have to be considered. They are
\eqal{	
T_{\mu\nu}\arrowvert_{\*\l}&=\Fraction{\th+1}{6}{\D}^{k-{3\/2}}
	{\*}_{m}{\D}{\g}_{[\bar{\mu}}{\S}^{\bar{m}}{\*}_{\nu]}\l\komma\\
T_{\mu m}\arrowvert_{\*\l}&={\D}^{k-1}{\*}_{n}{\D}
	\Bigl[(k-\Fraction{\th}{12}){\d}_{\bar{m}}{}^{\bar{n}}
	+\Fraction{\th+1}{6}{\S}_{\bar{m}}{\space}^{\bar{n}}\Bigr]
	{\*}_{\mu}{\l}\komma\\
T_{mn}\arrowvert_{\*\l}&=0\punkt
	}{\torfive} 
The $m$-component of the linearised equation of motion (\PsiEOM) becomes
$$
T_{m}\arrowvert_{\*\l}=
	-{\D}^{k-{5\/6}}{\*}_{n}{\D}
	\Bigl[(k-\Fraction{\th}{12}){\d}_{\bar{m}}{}^{\bar{n}}
	+\Fraction{\th+1}{6}{\S}_{\bar{m}}
	{}^{\bar{n}}\Bigr]{\g}^{\bar{\mu}}{\*}_{\mu}{\l}=0\punkt\eqn
$$
This equation gives immediately the Dirac equation,
$\g^{\mu}{\*}_{\mu}{\l}=0$, for any mode except the one corresponding
to the unbroken supersymmetry. 
Using the Dirac
equation the other component of the trace can be written as
$$
T_{\mu}\arrowvert_{\*\l}=
	{\D}^{k-{4\/3}}{\*}_{m}{\D}{\S}^{\bar{m}}
	{\*}_{\mu}{\l}(k+\Fraction{\th}{4}+\Fraction{1}{3})=0
\punkt\eqn
$$
Assuming that the broken supersymmetry must have the opposite
chirality compared to the unbroken one (as must be case if the zero-mode
is part of a $D=6$ tensor multiplet), we get from the equation above
that $k=-{7\/12}$. Thus, we have obtained the equation of
motion and the transversal behaviour of the Goldstone fermion living on 
the M5 brane. Performing the same calculations for the M2 brane 
we obtain $\th=+1$
and $k=-{1\/6}$ for the unbroken supersymmetry and $\th=-1$ and
$k=-{4\/6}$ for the broken one. The only difference in this calculation is that
$\th$ now denotes the 8-dimensional chirality of $\zeta$.

In order to do the same calculation for the D3 brane, we take $\zeta$ to have
positive 10-dimensional chirality which implies that the 4 and 6-dimensional
chiralities of $\zeta$ must be the same, denoted by $\th$. With the convention
that
\eq{\g^{\m\n\r\s}=i\e^{\m\n\r\s}\g^5\komma
}
we find that $\th=-1$
and $k=-{1\/8}$ for the unbroken supersymmetry while $\th=+1$ and $k=-{5\/8}$
for the broken one.

To check the obtained values of $k$ we now examine the supersymmetry algebra,
which we schematically write as
\eql{[\zeta Q,\zeta'Q]=\zeta\G^M\zeta'\sL_M+\ldots=\e^M\sL_M+\ldots}{\QQL}
(only the diffeomorphisms are written out in the RHS).
The unbroken supersymmetry must generate translations in the longitudinal
directions that have no $y$-dependence, giving that
\eq{\zeta\G^\m\zeta '=(\D^k\l)(\D^{-\a}\G^{\bar{\m}})(\D^k\l')}
must be independent of $y$. We thus get $k=\Fraction{\a}{2}$ for the unbroken
supersymmetry in agreement with our previous results.
In order to generate a translation in the transverse directions, which we have
seen must behave as $\e^m=\D^{-1}\phi^m$, we must commute a broken and an 
unbroken
supersymmetry generator, giving
\eq{\zeta\G^m\zeta '=(\D^k\l)(\D^{-\b}\G^{\bar{m}})(\D^{k'}\l')
=\D^{-1}\phi^m\komma}
where $k'$ denotes the exponent for the broken supersymmetry. 
We get $k'=\b-k-1$,
which also agrees with our previous results. Inclusion of the tensor gauge
transformations of the following two subsections
in the RHS of eq. (\QQL), thus relating the
transverse behaviour of the fermionic and tensorial modes, gives the
same result.

In all three cases under consideration, we find that the fermion zero-modes
are normalisable in the transverse directions, so that the effective action
reduces to that of a chiral spinor\footnote\dagger{For the membrane, chirality
refers to the internal Spin(8) indices.} in the longitudinal directions.
We also note that the assumption concerning the chirality of the fermion 
zero-modes made above is unnecessary, since the field strength obtained
from the other chirality, \ie, the conserved supersymmetry, is identically
zero. 

\subsection\VectorZM{The Vector Zero-modes}We will now see the first 
example of how broken tensor gauge symmetries give
rise to zero-modes. We know that there is a vector field living on the D3
brane and the question is now how to interpret this field as arising 
from broken
gauge symmetries. Since a vector zero-mode comes from a broken vector, or 1-form, gauge
parameter we must have a corresponding 2-form potential and 3-form field
strength in which the zero-modes live. Of course, $D=10$ supergravity contains
a 3-form field strength $H$ corresponding to 
the 2-form potential $B$ [\HoweWest]. 
It is important to note that these supergravity fields are complex. 
We now make a gauge transformation $\d B=d\L$ and make the Ansatz
$\L=\D^k A$, where $A$ is a constant 1-form which lies 
in the longitudinal
directions. The reason why we take $A$ to lie in the longitudinal directions 
is of course that we want to be able to integrate out the transversal 
dependence, thus obtaining an effective vector theory on the brane 
world-volume. We get
\eq{\d B=d\D^k \wdg A\punkt
}
We now let $A$ become $x$-dependent, which means that it is no longer 
a pure gauge
transformation. By computing $\d H$ and solving the equations of motion 
for the variation we will get the equations of motion and the transversal
behaviour for the zero-modes. We find
\eq{\d H=d(\d B)=-d\D^k\wdg F\komma
}
$F$ being the complex 2-form field strength on the brane, $F=dA$,
and when we look at the equations of motion [\HoweWest]
\eql{d{\star}H+iG\wdg H=0}\HvarEOM
and use $G$ from (\threesol), we get
\eql{\D d{\star_x}F\wdg{\star_y}d\D+(iF-k{\star_x}F)\wdg d\D\wdg{\star_y}d\D=0
\punkt}\FinalTEOM
The notation $\star_x$ and $\star_y$ implies dualisation with flat metrics
$\eta_{\m\n}$ and $\d_{mn}$. In the case of the longitudinal 2-form $F$,
this makes no difference from using the restriction of the actual metric. 

We now consider the two four-dimensional duality components of $F$, fulfilling
${\star_x}F=\pm iF$, separately. The two terms in eq. (\FinalTEOM) have 
different index structure and vanish independently.
The first term gives that
$dF=0$, which together with the relation ${\star_x}F=\pm iF$ is the equation of
motion for $F$ and hence for the zero-modes. The second term determines the
value of $k$ and we have $k=1$ for the positive sign and $k=-1$ for the
negative sign. Each duality component of $F$ contributes with two
modes and na\"\i vely we thus have twice the number of 
modes we wanted. However,
by requiring normalisability the positive sign is forbidden and we are left
with the desired number of zero-modes.

The (anti-)self-duality of $F$ reflects the self-duality 
property of the D3 brane
itself, and is connected to the fact that it forms a singlet under the 
SL(2;$\Z$) symmetry of type \II B. Since we use a formalism for the
supergravity where this symmetry is manifest, we do not obtain the real
vector potential of the Born--Infeld theory for the zero-modes, but instead
a complex one satisfying a complex self-duality. This ties up naturally
with the work of ref. [\CW], where the D3 brane was given an 
SL(2;$\Z$)-covariant formulation. The fields used there are identical to
the ones obtained here.

\subsection\TensorZM{The Tensor Zero-modes}Finally, we discuss the 
tensor modes living on the M5 brane. As
mentioned in the beginning of this section, these modes are related to
the breaking of the gauge symmetry of the background 3-form potential
$C$. Hence, we consider an infinitesimal gauge 
transformation
$\d C=d\L$ and make the Ansatz
$\L={\D}^{k}A$, where $A$ is a constant 2-form
which lies in the longitudinal directions. Along the lines of the
previous discussions we let $A$
become $x$-dependent and use the Ansatz to obtain the following expression
for the variation of the $H$-field
\eql{	h=\d{H}=F\wdg d\D^k\komma
	}{\varH} 
where $F=dA$. We now require
this expression to satisfy the equation of motion (\Heqv) to linear order,
\eql{	d{\star}h-H\wdg{h}=0\punkt
	}{\Heqvlin} 
We thus obtain
\eq{{\D}d{\star_x}F\wdg{{\star_y}d\D}
-(k{\star_x}F-F)\wdg d{\D}\wdg{{\star_y}d\D}=0\punkt
	}

We now consider the two 6-dimensional duality components of $F$ (fulfilling
${\star_x}F=\pm{F}$) separately. The first term in the equation above
gives immediately that $dF=0$, which together with the relation
${\star_x}F=\pm F$ is the equation of motion for the
$F$-field. The second term determines the constant $k$ and we have
$k=-1$ for the anti-self-dual part and $k=1$ for the self-dual part. Each
duality component of $F$ contributes with three modes and as in the
case of the vector zero-mode we na\"\i vely have twice as many zero-modes as we
wanted. Again, normalisability forbids one part of $F$, 
in this case the self-dual part.
Of course, the quadratic $H$ term in the action vanishes 
when an (anti-)self-dual field is inserted. This may be seen as a cancellation 
between the kinetic and potential parts of $L=K-V$. Demanding that the 
{\it energy} $E=K+V$ is finite per unit brane volume amounts to the na\"\i ve 
transverse normalisability condition on the mode function. 
It is also noteworthy
that the combination entering the action, namely the product of the two
chiralities, is not normalisable, so the self-dual component can
not serve as an auxiliary field. 
We finally note that the transversal behaviour for the normalisable
tensor zero-modes given in [\Kaplan] does not seem correct.

\subsection\Sumzero{Summary of Normalisable Zero-modes}We end this 
section by noting that we
have obtained the transverse behaviour and the equations of motion 
for all the zero-modes 
living on the M2, M5 and D3 brane by following a
common procedure. 
The normalisable zero-modes we found are most conveniently summarised
in terms of the gauge parameters containing the moduli:
$$
\matrix{&&&\hbox{\bf M2}&\hbox{\bf M5}&\hbox{\bf D3}\cr
\hbox{Diffeomorphisms:}\hfill&\e^m\hfill&=\hfill&\D^{-1}\phi^m\hfill
&\D^{-1}\phi^m\hfill&\D^{-1}\phi^m\hfill\cr
\hbox{Local supersymmetry:}\hfill&\zeta\hfill&=\hfill&\D^{-{2\/3}}\l_-\hfill
&\D^{-{7\/12}}\l_+\hfill&\D^{-{5\/8}}\l_+\hfill\cr
\hbox{Tensor gauge symmetry:}\hfill&\L\hfill&=\hfill&&\D^{-1}A\hfill
&\D^{-1}A\hfill\cr
&&&&({\star}F=F)\hfill&(i{\star}F=F)\hfill\cr}\Eqn\Summary
$$

As already mentioned, it is straightforward to check explicitly that the 
collective coordinates, 
\ie, the fields on the branes, form multiplets under the unbroken
supersymmetries generated by the Killing spinors of section \FermZM,
thus providing a further check that the modes in eq. (\Summary) are correct.

We note that the two distinguished fermionic modes, namely the Killing
spinor and the fermionic zero-mode, in addition
to having opposite chiralities, carry different dependence on the transverse
coordinates. It does not make sense, except asymptotically, 
to think of the two as making up
a non-chiral spinor of broken + unbroken supersymmetry. The zero-modes
are in fact just one out of an infinite number of supersymmetries broken
by the brane solution.

The vector modes of the D3 brane and the tensor modes of the M5 brane
follow a very similar pattern, where only one out of two duality components
is allowed by normalisability (finite energy condition).

In the Ans\"atze used for finding the transverse behaviour of the 
collective coordinates, we have used the constant mode on the 
transverse spheres. In principle, a general Ansatz would contain also
higher Kaluza--Klein modes, but that discussion was postponed for 
simplicity. It is straightforward, and we will not go into details here, 
to show that such higher modes
will not contain zero-modes---they will lead to massive fields on the brane.
Another assumption, motivated by the knowledge of the presence of vector
or tensor modes, was the index structure of the tensor gauge transformations, 
\ie, that the gauge parameters should carry only longitudinal indices
(and corresponding statements for the other modes).
This assumption is also straightforwardly verified by considering
a more general Ansatz---the self-duality is the essential property that
enables us to obtain massless modes by a cancellation between the two 
terms, originating from the kinetic term and Chern--Simons term,
in the tensor equations of motion (\HvarEOM) or (\Heqvlin).
 
\section\bps{Charges of the Excited M5 Brane}The zero-modes of the M5 brane
form a multiplet under the unbroken supersymmetry generated by the
Killing spinors that make eq. (\fivepot) vanish, \ie, under a 6-dimensional
(2,0) supersymmetry algebra. The amount of supersymmetry of the solution
equals half the number of Killing spinors of the asymptotic Minkowski 
space---the extremal brane solutions are half-supersymmetric BPS 
configurations. The central charge in the 11-dimensional supersymmetry
algebra is a 5-form, the magnetic charge of the M5 brane. This algebra
is obtained by anticommuting the unbroken and broken supersymmetries of
section \FermZM\ in the asymptotic Minkowski region and using the background
value of the tensor field. 

Here we want to discuss briefly the corresponding situation when the
tensorial zero-modes are excited. 
The first thing to observe is that once the tensor mode is turned on,
the M5 brane no longer carries only magnetic charge, but also electric. 
The electric charge is measured by the flow out of a closed 7-surface. 
If this hypersurface is the contractible boundary of an 8-volume $\M_8$,
the electric charge vanishes:
$$
e=\int_{\dM_8}({*}H-\half H\wedge C)=\int_{\M_8}(d{*}H-\half H\wedge H)=0
\komma\Eqn\ElCharge
$$
due to the equation of motion for $H_{(4)}$ (or equivalently, 
the Bianchi identity for $H_{(7)}$).
If however $\M_8$ intersects with the horizon $\S_5$ of the M5 brane, $\r=0$,
its boundary can not be freely contracted. Without deforming the
intersection with $\S_5$, one may contract to the product of the intersection
and a small 5-ball centered around each point in the intersection:
$\M_8\rightarrow\M'_8=(\M_8\cap \S_5)\times B^5(\e)$. The electric
charge may now be calculated as the integral over $\dM'_8$. Using the
explicit solution of section \TensorZM, 
the only contribution comes from the first
term in the integral in eq. (\ElCharge) over 
$(\M_8\cap \S_5)\times S^4(\e)$, and the result is 
$$
e=q\int_{\M_8\cap \Sigma_5}F\komma\Eqn\ElChargeTwo
$$
where $q=\int_{S^4}H=8\pi^2R^3$ is the magnetic charge of the 5-brane.
The corresponding statement is true for the D3 brane,
where an excitation of the Born--Infeld field carries charge with respect
to the NS-NS and RR 2-form potentials. A D${}_p$-brane
generically intersects the 8-dimensional hypersurface defining the charge in a
$(p-1)$-dimensional hypersurface, and the charge will have to be expressed
as an integral $e\sim\int_{\M_8\cap \Sigma_p}F_{(p-1)}$.
The fields $F_{(p-1)}$ are those naturally obtained in 
the formalism of ref. [\CW].

The expression for the electric charge (\ElChargeTwo) is a topological
quantity on the brane. The 3-dimensional manifold may enclose stringlike
objects, the self-dual string solitons of ref. [\StringSoliton], 
and the integral measures the string charge.
 
The electric charge of the branes with excited tensors/vectors parallels
the situation for monopoles in field theory, where momentum in the
fourth direction of the moduli space is identified as electric charge.
The analogy here is the field strength on the brane. As for the monopoles,
the electric charge in equations (\ElCharge) and (\ElChargeTwo) is
the classical expression---charge quantisation is not seen at this level.

It is known from existing formulations of the dynamics of branes
with vector or tensor degrees of freedom 
[\KappaBranes,\CT,\CW,\CNS,\FivebraneLagr] 
that the brane actions
are $\k$-symmetric, \ie, there exists a projection of the target space
spinor coordinates that are purely gauge degrees of freedom.
For the M5 brane this projection looks like [\CNS]
$P={1\/2}(\id\pm\G)$, where (barring numerical factors)
$$
\G\sim{\e^{\m_1\ldots\m_6}\/\sqrt{-g}}(\G_{\m_1\ldots\m_6}
	+F_{\m_1\m_2\m_3}\G_{\m_4\m_5\m_6})\komma\eqn
$$
The first term defines chirality with respect to the spinor decomposition
(\gammasplitgen) and are related to the magnetic charge of the M5 brane,
or, more generally, to (electric) charge with respect to a potential 
$C_{(p+1)}$, while the second term is related to the electric charge
(\ElChargeTwo), and corresponds to a 2-form extension in the 11-dimensional
supersymmetry algebra.
In reference [\SorokinTownsend], extensions of the 
11-dimensional supersymmetry algebra
were analysed from the point of view of the M5 brane, and it was shown
how the 2-form extension is related to the self-dual tensor. Here, we have
added a more direct ingredient to this interpretation, namely the resulting
charges obtained by using the explicit form of the tensor field $H$ when the
self-dual tensor is excited.

The above analysis is presented here only to linear
order in $F$, and we refer to a forthcoming publication [\NextPaper] for 
the case of finite field strengths.

\section\sc{Discussion}We have presented a principle, intimately 
connected to gauge symmetries, by which the
zero-modes around a brane solution in supergravity can be found and
explicitly constructed. The procedure has been carried through in detail
for the membrane and 5-brane in eleven dimensions, as well as for the D3 brane
of type \II B in ten dimensions. 

The nature of the construction emphasises the way the
zero-modes arise as Goldstone modes of a broken global symmetry.
We would like to stress again that the relevant symmetries of the 
(supergravity) theories in question are gauge symmetries: 
reparametrisations, local supersymmetry and gauge symmetry of the
tensor fields. The zero-modes building the supersymmetric field theories
on the branes arise as Goldstone modes for breaking of certain modes of
these symmetries which are large gauge transformations, and are as such
global rather than local symmetries in the given backgrounds. The property
of the brane configurations permitting such transformations is the fact
that they contain different unconnected asymptotic regions (the asymptotic
Minkowski region far from the brane and the near-horizon AdS region), and
the gauge parameters may take different values in the two regions. It is 
easily seen that the relevant modes have exactly this property. This 
hinges on the fact that the transverse behaviour of the collective
coordinates carry {\it negative} powers of the harmonic functions $\D$,
so that the difference from horizon to infinity is well defined and
finite: $(\lim_{\r\rightarrow\infty}-\lim_{\r\rightarrow0})\D^{-p}=1$.

The tensor or vector modes have sometimes been considered as more 
mysterious than the other ones, especially concerning their Goldstone
properties [\Bagger] (maybe as a result of picturing the Goldstone mechanism
as connected to asymptotic isometries instead of large gauge transformations). 
The present analysis, in contrast, treats all modes 
on equal footing. We should maybe remark that, although the popular
picture of \eg\ the scalar modes as Goldstone modes for the breaking of
translational symmetry is appealing, the notion of translational symmetry
in a gravity theory is somewhat suspect. The relevant symmetry is a
large reparametrisation, having different asymptotic values far from
the brane and near the horizon. It does not correspond to a rigid shift of the
transverse coordinates in the solutions of section \prel.

It is satisfying to see that the tensor or vector fields that arise
are exactly those that appeared in refs. [\CW,\CNS], where they arose 
as fields on the branes, having the natural couplings to background fields
and reflecting the symmetries of the background theory.

The entire construction is performed at linear level around infinite flat
branes, which means that some aspects connected to 
non-linearities---Born--Infeld dynamics of D-branes and the corresponding
non-linear dynamics in eleven dimensions---are not seen. Although these
represent higher-derivative terms in an action, and therefore seem 
irrelevant for a discussion of the low-energy behaviour, these excitations
are such that the BPS property is exactly preserved. States found in a
quantum field theoretic treatment of the full non-linear theory are therefore
reliable; the concept of low energy should really be replaced by preservation
of the BPS property. 
We have recently noted that it is possible to solve the coupled system
of equations of motion for the gravity and tensor field to all orders,
which results in a brane solution with (constant) finite tensor
field strength. This will be reported in ref. [\NextPaper].

\appendix{Conventions}We use the conventions that
\eq{	\eta_{\m\n}=(-1\; +1 \;+1 \; \ldots\; +1)\komma
	} 
and that the Levi-Civita tensor density $\e$ with downstairs
indices is defined to be +1.
 
Our convention for dualisation of a $p$-form in $D$ dimensions is
$$
{\star}(dx^{M_1}\wedge dx^{M_2}\wedge \ldots
\wedge dx^{M_p})={\sqrt{|g|}\/(D-p)!}\e^{M_1\ldots M_p}{}_{M_{p+1}\ldots M_D}
dx^{M_{p+1}}\wedge \ldots \wedge dx^{M_D}\komma\eqn
$$
which translated to acting on the tensor components reads
$$
({\star}\W)_{M_1\ldots M_{D-p}}={{\sqrt{|g|}\/p!}\W_{N_1\ldots N_p}
\e^{N_1\ldots N_p}{}_{M_1\ldots M_{D-p}}}
\komma\eqn
$$
with the convention that
$$
\W_{(p)}={1\/p!}dx^{M_1}\wedge dx^{M_2}\wedge \ldots
\wedge dx^{M_p} \W_{M_1\ldots M_p}\punkt\eqn
$$

\acknowledgements 
The authors would like to thank M\aa ns Henningson 
for discussions.
\refout

\end